\documentclass[camera,letterpaper,nomarginnotes,nonarrowgutter]{jpaper}
\usepackage{cite}
\usepackage{amsmath,amssymb,amsfonts}
\usepackage{algorithmic}
\usepackage{graphicx}
\usepackage{textcomp}
\usepackage{balance}
\usepackage{fancyhdr}
\usepackage{glossaries}
\usepackage{siunitx} 
\usepackage{booktabs} 
\usepackage{multirow}
\usepackage{xcolor}
\usepackage{enumerate}
\usepackage{hyperref}

\def\BibTeX{{\rm B\kern-.05em{\sc i\kern-.025em b}\kern-.08em
    T\kern-.1667em\lower.7ex\hbox{E}\kern-.125emX}}

\def\UrlBreaks{\do\/\do-\/\do.\/\do:}

\expandafter\def\expandafter\UrlBreaks\expandafter{\UrlBreaks
  \do\a\do\b\do\c\do\d\do\e\do\f\do\g\do\h\do\i\do\j
  \do\k\do\l\do\m\do\n\do\o\do\p\do\q\do\r\do\s\do\t
  \do\u\do\v\do\w\do\x\do\y\do\z\do\A\do\B\do\C\do\D
  \do\E\do\F\do\G\do\H\do\I\do\J\do\K\do\L\do\M\do\N
  \do\O\do\P\do\Q\do\R\do\S\do\T\do\U\do\V\do\W\do\X
  \do\Y\do\Z}

\definecolor{amber}{rgb}{1.0, 0.49, 0.0}
\definecolor{awesome}{rgb}{1.0, 0.13, 0.32}
\definecolor{azure}{rgb}{0.0, 0.5, 1.0}
\definecolor{dollarbill}{rgb}{0.52,0.73,0.4}
\definecolor{moegi}{rgb}{0.357, 0.537, 0.188}
\definecolor{burgundy}{rgb}{0.5, 0.0, 0.13}
\definecolor{ballblue}{rgb}{0.13, 0.67, 0.8}
\definecolor{ups-truck}{rgb}{0.53, 0.28, 0.21}
\definecolor{airforceblue}{rgb}{0.36, 0.54, 0.66}
\definecolor{cadmiumgreen}{rgb}{0.0, 0.42, 0.24}
\definecolor{darkcyan}{rgb}{0.0, 0.55, 0.55}
\definecolor{caribbeangreen}{rgb}{0.0, 0.8, 0.6}
\definecolor{flamingopink}{rgb}{0.99, 0.56, 0.67}
\definecolor{jazzberryjam}{rgb}{0.65, 0.04, 0.37}
\definecolor{mediumpersianblue}{rgb}{0.0, 0.4, 0.65}
\definecolor{coolblack}{rgb}{0.0, 0.18, 0.39}
\definecolor{bleudefrance}{rgb}{0.19, 0.55, 0.91}
\definecolor{ao}{rgb}{0.0, 0.0, 1.0}
\definecolor{babyblueeyes}{rgb}{0.63, 0.79, 0.95}
\definecolor{darkwarmgray}{rgb}{0.2,0,0}

\newcommand{\mechanism}{SpyHammer}

\newacronym{hcfirst}{\emph{HC\textsubscript{first}}}{the minimum hammer count at which the first bit error is observed}
\newacronym{ber}{$BER$}{Bit Errors per Row}
\newacronym{wcdp}{$WCDP$}{worst-case data pattern}

\newcounter{take}
\newcommand\take[1]{%
   \refstepcounter{take}
  \vspace{0.5em}
  \noindent
  \begin{tabular}{|p{0.95\linewidth}|}
       \hline
       \textbf{{Takeaway \thetake}.} \emph{{#1}}\\
       \hline 
  \end{tabular}
  \vspace{-0.5em}
  \addtocounter{table}{-1} 
}

\newcommand{\obslbl}[0]{Observation}

\newcounter{obs}
\newcommand\observation[1]{%
   \refstepcounter{obs}
   \noindent
   \colorbox{gray!20}{\textbf{\obslbl{} \theobs.}} \emph{#1}
}

\newif\ifcameraready
\camerareadytrue

\newif\ifrevision
\revisionfalse

\newif\ifsubmission
\submissiontrue

\newcommand{\loiscomment}[1]{\textcolor{red}{\textbf{[@lois:} #1\textbf{]}}}

\newcommand{\atbc}[1]{\textcolor{azure}{\textbf{Atb:}#1}}
\newcommand{\agy}[1]{\textcolor{pink}{#1}}
\newcommand{\lois}[1]{\textcolor{ballblue}{#1}}
\newcommand{\minesh}[1]{\textcolor{black}{#1}}
\newcommand{\agycomment}[1]{\textcolor{ballblue}{\textbf{[@gy:} #1\textbf{]}}}

\newcommand{\onur}[1]{\textcolor{black}{#1}}

\ifsubmission
\renewcommand{\agy}[1]{\textcolor{black}{#1}}
\renewcommand{\agycomment}[1]{}
\renewcommand{\loiscomment}[1]{}
\renewcommand{\atbc}[1]{}
\fi
\ifrevision

\fi
\ifcameraready

\renewcommand{\agy}[1]{\textcolor{black}{#1}}
\renewcommand{\lois}[1]{\textcolor{black}{#1}}
\fi

 \newcommand{\affilETH}{$^1$}
 \newcommand{\affilMunich}{$^3$}
 \newcommand{\affilConnecticut}{$^{4}$}
 \newcommand{\affilMunichConnecticut}{$^{3,4}$}
 
 \newcommand{\affilCESGA}{$^2$}
 \newcommand{\affilETHCESGA}{$^{1,2}$}

 \author{
 {Lois Orosa\affilETHCESGA{}}\qquad%
 {Ulrich R{\"u}hrmair\affilMunichConnecticut{}}\qquad
 {A. Giray Ya\u{g}l{\i}k\c{c}{\i}\affilETH{}}\qquad%
 {Haocong Luo\affilETH{}}\qquad
 {Ataberk Olgun\affilETH{}} \\
 {Patrick Jattke\affilETH{}}\quad 
 {Minesh Patel\affilETH{}}\qquad
 {Jeremie Kim\affilETH{}}\qquad 
 {Kaveh Razavi\affilETH{}}\qquad
 {Onur Mutlu\affilETH{}}\\
  \qquad \\
  \affilETH\emph{ETH Z{\"u}rich}\qquad
  \affilCESGA\emph{Galicia Supercomputing Center (CESGA)}\qquad \\
  \affilMunich\emph{LMU M{\"u}nchen}\qquad 
  \affilConnecticut\emph{University of Connecticut}\qquad
 }

\title{SpyHammer: \\Understanding and Exploiting RowHammer \\ under Fine-Grained Temperature Variations}

\begin{document}
\bstctlcite{IEEEexample:BSTcontrol}

\maketitle

\begin{abstract}
RowHammer is a DRAM vulnerability that can cause bit errors in a victim DRAM row \minesh{solely} by accessing its neighboring DRAM rows at a high-enough rate. Recent studies demonstrate that new DRAM devices are becoming increasingly vulnerable to RowHammer, and many works demonstrate system-level attacks for privilege escalation or information leakage. In this work, we perform the first rigorous fine-grained characterization and analysis of the correlation between RowHammer and temperature. We show that RowHammer is very sensitive to temperature variations, even if the variations are very small (e.g., $\pm$\SI{1}{\celsius}). We leverage two key observations from our analysis to \minesh{spy on DRAM temperature}:
1)~RowHammer-induced bit error rate consistently increases (or decreases) as the temperature increases, and 2)~some DRAM cells that are vulnerable to RowHammer \minesh{exhibit} bit errors only at a particular temperature.
Based on these observations, we propose a new RowHammer attack, called \mechanism{}, that spies on the temperature of \minesh{DRAM on} critical systems such as industrial production lines, vehicles, and medical systems. \mechanism{} is the first practical attack that can spy on DRAM temperature. Our evaluation in a controlled environment shows that \mechanism{} can infer the temperature of the victim DRAM modules with an error of less than $\pm$\SI{2.5}{\celsius} at the $90^{\mathrm{th}}$ percentile of all tested temperatures, for 12 real DRAM modules (120 DRAM chips) from \onur{four} main manufacturers.


\end{abstract}

\section{Introduction}
RowHammer is a DRAM vulnerability where a DRAM cell experiences a bitflip when its nearby cells are rapidly and frequently accessed~\cite{kim2014flipping}.
Recent works~\cite{kim2020revisiting, frigo2020trrespass} demonstrate that modern DDR4 DRAM devices are more vulnerable to RowHammer than their predecessor DDR3 devices, suggesting that RowHammer is an important DRAM \minesh{system design concern} that is becoming increasingly severe as DRAM manufacturing technology nodes scale down.
Using RowHammer, many works demonstrate attacks that escalate privilege at system-level, leak secret information, and manipulate critical application outputs~\cite{seaborn2015exploiting, van2016drammer, gruss2016rowhammer, razavi2016flip, pessl2016drama, xiao2016one, bosman2016dedup, bhattacharya2016curious, qiao2016new, jang2017sgx, aga2017good, mutlu2017rowhammer, tatar2018defeating, gruss2018another, lipp2018nethammer, van2018guardion, frigo2018grand, cojocar2019eccploit,  ji2019pinpoint, mutlu2019rowhammer, hong2019terminal, kwong2020rambleed, frigo2020trrespass, cojocar2020rowhammer, weissman2020jackhammer, zhang2020pthammer, rowhammergithub, yao2020deephammer, jattke_blacksmith_2022,cohen2022hammerscope,tobah2022spechammer}.

\lois{We perform the first rigorous fine-grained characterization and analysis of the correlation between RowHammer and temperature, which lead to eight insightful observations and three key takeaways.}
\lois{Based on our observations and takeaways,} we demonstrate \mechanism{}, a new attack that uses RowHammer to spy on DRAM temperature with high accuracy. Our attack can be performed with minimal knowledge of the target computing system and can be used to compromise the security, confidentiality \minesh{and privacy} of critical systems that use DRAM. \minesh{\mechanism{}} can compromise a victim computing system to achieve two goals. First, it can identify the utilization of a computer system, as the compute and memory intensity of a workload can change the temperature of the system. For example, an attacker can use \mechanism{} to infer \minesh{when} a server is at its peak \minesh{utilization} by spying \minesh{on} its temperature. Second, \minesh{\mechanism{}} can measure the ambient temperature, which may convey information about the state of a larger system that contains the target computing system (e.g., a car, a drone, or an industrial manufacturing machinery). For example, the temperature of a car's engine may rise if the engine is operating at high revolutions per minute. 

\minesh{\mechanism{} not only compromises security and confidentiality, but also privacy}. For example, by spying the temperature of a house (or different rooms of a house), an attacker can infer the habits of the person(s) living in that house. Tracking the temperature could give information about at which times the person(s) \agy{leave} or enter \agy{a room in} the house.



\minesh{\mechanism{} leverages} two key observations about RowHammer to spy on DRAM temperature: 1)~RowHammer-induced bit error rate consistently increases (or decreases) when temperature increases, and 2)~some DRAM cells that are vulnerable to RowHammer experience bit errors only at a specific temperature. Using these observations, \minesh{\mechanism{}} infers the temperature of DRAM chips by only characterizing DRAM cells that exhibit RowHammer-induced bit errors in the address space of the attacker without requiring any hardware or system software modifications. We propose two variants of the \mechanism{} attack, each with a different threat model.

The first variant of \mechanism{} can identify \emph{relative temperature changes}\minesh{, and it does not require prior physical access to or knowledge about the victim DRAM module}. 
\minesh{We observe that the correlation between \gls{ber} and temperature follows a similar trend in different DRAM modules of the same model and manufacturing date.}
We use this observation to spy on relative temperature changes. \minesh{The key idea is to infer the model and manufacturing date of the victim DRAM module, and use a module with the same characteristics \minesh{(to which the attacker has physical access and can control the operating temperature of)} to infer the correlation between \gls{ber} and temperature of the victim DRAM module}. Since the attacker has \emph{no} prior information about the victim DRAM module, the attacker must reverse engineer the victim DRAM module using remote RowHammer-based techniques~\cite{lipp2018nethammer,tatar2018throwhammer}. To estimate the temperature of the victim DRAM module, we propose to build a polynomial regression model using a DRAM module \minesh{that has similar characteristics} as the victim DRAM module.

The second variant of \mechanism{} spies on \emph{absolute temperatures}, which requires characterizing the victim DRAM module before the attack. The key idea is to build an accurate polynomial regression model using the characterization data of the victim DRAM module. This model is then used in the attack to \agy{accurately} infer the \agy{victim DRAM module's} temperature. 

Reliably monitoring the \gls{ber} of a DRAM module requires the attacker to hammer a large region of memory, which might increase the complexity of the attack. To reduce the \minesh{number of DRAM accesses to the victim DRAM module}, we propose \minesh{a \mechanism{}} optimization that leverages the observation that some DRAM cells \agy{experience bitflips} only at one particular temperature. \minesh{We call these cells \emph{canary cells}}.\footnote{In all possible temperature points within a temperature range (given a particular temperature resolution), a canary cell \agy{experiences a bitflip} at one and only one temperature point.}

The enrollment phase identifies the canary cells of the DRAM module. In this process, the cells that flip at only one temperature are added to the canary cell set. After the enrollment phase, an attacker can estimate the temperature of the victim DRAM by monitoring only a few selected canary cells, which reduces the number of memory accesses required to perform the attack.

To evaluate \mechanism{}, we perform an extensive and thorough DRAM RowHammer characterization on 12 real DRAM modules (120 DRAM chips) using a temperature resolution of \SI{1}{\celsius} in a controlled environment. Our results show that our methodology can infer 1)~absolute temperatures (with prior characterization of the victim DRAM module) with an error of $\pm$\SI{2.5}{\celsius}, and 2)~relative temperature changes (without prior characterization of the victim DRAM module) with an error of $\pm$\SI{3.5}{\celsius}, for all 12 DRAM modules we test, at the $90^{\mathrm{th}}$ percentile of tested temperature points (i.e., from \SI{50}{\celsius} to \SI{95}{\celsius}, with \SI{1}{\celsius} step size). 


We make the following main contributions:
\begin{itemize}

    \item We perform the first rigorous \emph{fine-grained} characterization and analysis of the correlation between RowHammer and temperature using 12 real DDR4 DRAM modules (120 DRAM chips).
    \item We show that RowHammer is very sensitive to temperature variations, even if the variations are very small (e.g., $\pm$\SI{1}{\celsius}).
    \item We propose \mechanism{}, the first \minesh{RowHammer} attack that can spy on DRAM temperature \agy{\emph{without}} any modification to the victim system. \mechanism{} uses only the \agy{attacker's memory space \lois{to perform the attack} (i.e., it does \emph{not} corrupt the victim's} memory space).
    \item We propose two variants of \mechanism{}: 1)~a variant that can spy on relative temperature changes \agy{\emph{without} any prior information about or changes to}
    the victim DRAM module, and 2) a variant that can spy on absolute temperature changes when the attacker has physical access to the victim DRAM module before deploying the attack.


    


    
    \item We perform a detailed study of the accuracy of the two \mechanism{} variants, which shows that an attacker can spy with a maximum error of 1)~$\pm$\SI{2.5}{\celsius} on absolute temperature values, and 2)~$\pm$\SI{3.5}{\celsius} on relative temperature changes, in \emph{all} 12 DRAM modules (120 DRAM chips) from the four \agy{major} manufacturers we test.\footnote{At the $90^{\mathrm{th}}$ percentile of tested temperature points} 
\end{itemize}
\section{Background}

We provide a brief introduction to DRAM organization and RowHammer vulnerability. For more detailed background, we refer the reader to prior works~\cite{liu2012raidr, liu2013experimental, keeth2001dram, mutlu2007stall, moscibroda2007memory, mutlu2008parbs, kim2010atlas, subramanian2014bliss, salp, kim2014flipping, qureshi2015avatar, hassan2016chargecache, chang2016understanding, lee2017design,  chang2017understanding,  patel2017reaper,kim2018dram, kim2020revisiting, hassan2019crow, frigo2020trrespass, chang2014improving, chang2016low, vampire2018ghose, hassan2017softmc, khan2016parbor, khan2016case, khan2014efficacy, seshadri2015gather, seshadri2017ambit, kim2018solar, kim2019d, patel2019understanding, patel2020beer, lee2013tiered, lee2015decoupled, seshadri2013rowclone, luo2020clrdram, seshadri2019dram, wang2020figaro,orosa2021codic,wang2018reducing,ipek2008self,zhang2014half,luo2023rowpress,ghose2018your,li2017drisa}.

\subsection{DRAM Organization}

The memory controller communicates with DRAM modules over one or more DRAM channels. Each module contains a set of DRAM chips that operate in lockstep. The DRAM cells within a DRAM chip are organized hierarchically. A DRAM chip comprises multiple DRAM banks that can operate independently. DRAM cells in a DRAM bank are laid out in a two-dimensional structure of rows and columns. Each DRAM cell on a DRAM row is connected to a common wordline via access transistors. A bitline connects a column of DRAM cells to a DRAM sense amplifier to access data. 

%

Accesses to DRAM devices are typically performed in cache block granularity (64-bytes) in contemporary systems. An access to a DRAM cache block works in three steps. First, the memory controller sends an ACT command to activate a specific row within a DRAM bank, which prepares the row for a columns access (i.e., copies the row to the sense amplifiers). Second, the memory controller sends a READ (WRITE) command to read (write) a column in the row. Third, once all operations to the active row are completed, the memory controller sends a PRE command that closes the row and prepares the DRAM bank to open a new DRAM row (i.e., it precharges the bank).

\subsection{RowHammer}
\label{sec:back_rowhammer}
Modern DRAM devices are subject to disturbance failures caused by high frequency accesses (i.e., hammer) to DRAM rows (i.e., aggressor rows) that result in bitflips in physically nearby rows that are not being accessed (i.e., victim rows). This phenomenon is referred to as RowHammer~\cite{kim2014flipping,kim2020revisiting,orosa2021Deeper,yauglikcci2022understanding,mutlu2023fundamentally,olgun2023experimental,mutlu2017rowhammer,mutlu2019rowhammer}. RowHammer-induced bitflips are exacerbated as DRAM technology nodes shrink and DRAM cells come closer to each other. This results in newer, higher-density DRAM chips to become more vulnerable to RowHammer~\cite{kim2020revisiting} and other read disturbance effects \cite{luo2023rowpress}. These bitflips
manifest after a row’s activation count reaches a certain threshold value within a refresh window (usually denoted as MAC~\cite{jedec2017ddr4} or $HC_{first}$~\cite{kim2020revisiting}).

Prior works devise many different \minesh{RowHammer-based} attacks, such as denial of service~\cite{gruss2018another,lipp2018nethammer}, privilege escalation~\cite{gruss2018another,lipp2018nethammer,gruss2016rowhammer,ji2019pinpoint,razavi2016flip,seaborn2015exploiting,van2016drammer,xiao2016one,jattke_blacksmith_2022,zhang2022implicit,kogler2022half}, secret data leakage~\cite{kwong2020rambleed,cohen2022hammerscope,tobah2022spechammer}, manipulation of the application correctness~\cite{hong2019terminal,yao2020deephammer} or private key recovery~\cite{fahr2022frodo,mus2023jolt}.
%
A subset of these attacks require no physical access to a victim computing system; for example, attacks leveraging RDMA~\cite{tatar2018throwhammer} or attacks in JavaScript programs~\cite{gruss2016rowhammer}.



\section{Methodology}
\label{sec:methodology}

In order to thoroughly characterize the correlation between RowHammer and temperature and analyze the potential of the \mechanism{} attack, we use an FPGA-based infrastructure that allows us to avoid uncontrolled interference in the system that might skew the results and lead to wrong insights and conclusions. 
%
To perform a \mechanism{} attack on a real commodity computer system, we can use the methodology proposed in previous works~\cite{frigo2020trrespass,jattke_blacksmith_2022,hassan2021uncovering} (not demonstrated in this paper).

\subsection{Testing Infrastructure}


We experimentally study DDR4  DRAM chips across a wide range of temperatures. We use the DRAM Bender framework~\cite{olgun2023dram,drambendergithub}, which supports DDR4 modules, and a highly accurate temperature controller infrastructure.

\subsubsection{DRAM Bender}

\begin{figure}[h] \centering
    \includegraphics[width=0.95\linewidth]{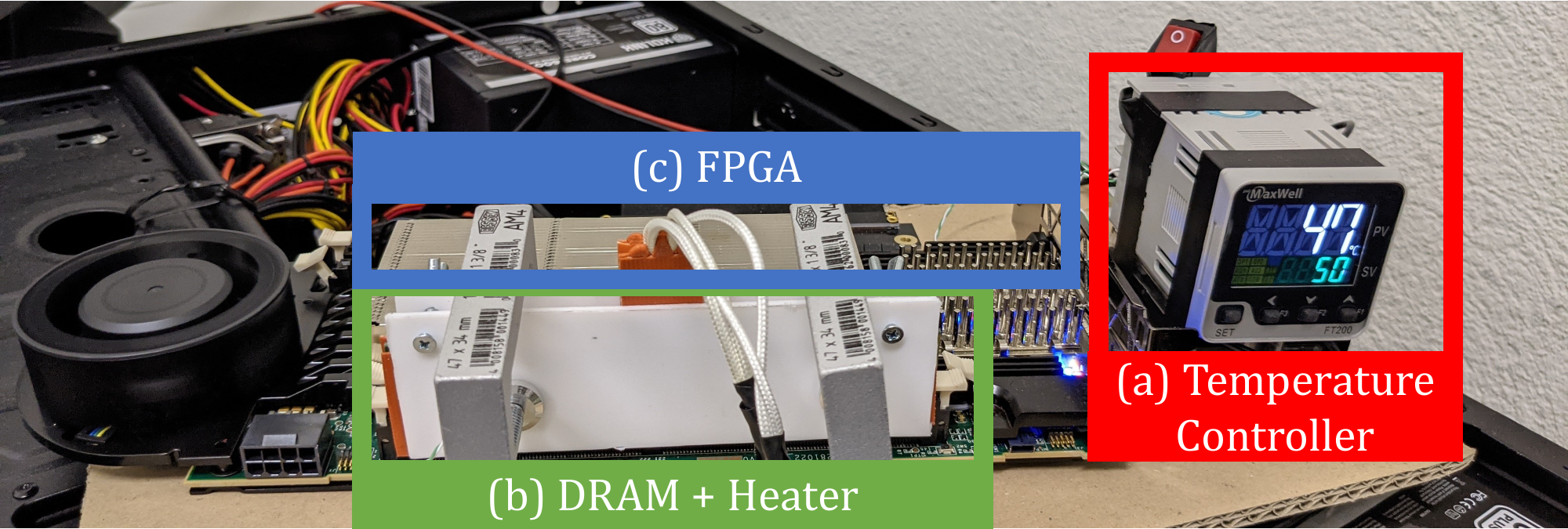}
    \caption{DRAM Bender Infrastructure: (a)~temperature controller, (b)~DRAM module clamped with heater pads, and (c)~FPGA board programmed with DRAM Bender}~\cite{olgun2023dram}.
    \label{fig:infrastructure}
\end{figure}

Figure~\ref{fig:infrastructure} shows the DRAM Bender setup for testing DDR4 DRAM modules. We use the Xilinx Alveo U200~\cite{alveo} FPGA board in all of our tests.

We use an FPGA board with DRAM Bender (Figure~\ref{fig:infrastructure}c) to perform all our RowHammer tests. We monitor and adjust the temperature of DRAM chips under test with a temperature controller (Figure~\ref{fig:infrastructure}a). This infrastructure provides \agy{us with} fine-grained control over the timing between DRAM commands. We enforce all timing parameters defined by JEDEC~\cite{jedec2017ddr4} to \lois{ensure} reliable operation. 

\subsubsection{Temperature Controller} \label{subsec:temp_controller}
To regulate the temperature in DRAM modules, we use silicone rubber heaters pressed to both sides of the DDR4 module (Figure~\ref{fig:infrastructure}c). To reduce the heat leakage, we apply two layers of insulation around the DRAM module under test and the heater pads: 1) a layer of reflective aluminum sheets covering the DRAM and the heater pads and 2) a layer of insulation sheets made of PTFE, a heat-resistant material.  To measure the actual temperature of DRAM chips, we use a thermocouple, which we place between the rubber heaters and the DDR4 chips. We connect the heater pads and the thermocouple to a Maxwell FT200 temperature controller (Figure~\ref{fig:infrastructure}a), which keeps the temperature stable by implementing a closed-loop PID controller. Our host machine communicates with the temperature 
controller via an RS485 channel. Using this feature, we build custom software that enables us to automate the management of the temperature and integrate it into our testing infrastructure. In our tests using this infrastructure, we measure temperature with an accuracy of $\pm$\SI{0.1}{\celsius}.


\subsection{Testing Methodology}

\vspace{0.2cm}
\noindent\textbf{Disabling Sources of Interference.} 
We disable all DRAM self-regulation events except the calibration signals, such as ZQ, for signal integrity so that we ensure that the observed errors are solely caused by RowHammer. We also make sure that our tests finish before retention errors manifest.

To the best of our knowledge, we also disable all DRAM-level (e.g., TRR~\cite{jedec2017ddr4}) and system-level RowHammer mitigation mechanisms (e.g., pTRR~\cite{aichinger2015ddr}) along with all forms of rank-level error-correction codes (ECC), which could obscure RowHammer bitflips.
Based on the prior work's observations~\cite{frigo2020trrespass, kim2020revisiting}, on-DRAM-die RowHammer mitigation mechanisms (i.e., TRR) take action when the DRAM services a refresh (REF) command. The DRAM modules we test do not implement error correction internally.

\vspace{0.2cm}
\noindent\textbf{RowHammer Access Sequence.} We use a common access sequence used in previous works \cite{kim2020revisiting,orosa2021Deeper, kim2014flipping} as the worst-case access pattern, in which we 1)~hammer the two rows that are adjacent to the victim row (i.e., aggressor rows), and 2)~access the aggressor rows as frequently as possible. 
In our tests, we perform a double-sided RowHammer attack~\cite{kim2014flipping, kim2020revisiting}.  

\vspace{0.2cm}
\noindent\textbf{Data Pattern.} We conduct our experiments on a DRAM module by using the module's \gls{wcdp}.
We identify the \gls{wcdp} as the pattern that experiences the largest number of bitflips among seven different data patterns that prior research on DRAM characterization uses~\cite{khan2014efficacy,liu2013experimental,patel2017reaper, kim2020revisiting, chang2016understanding, chang2017understanding, lee2017design}, presented in Table~\ref{tab:data_patterns}: 
{colstripe, checkered, rowstripe{, and} random {(we also test the {complements} of the first three)}.} 
For each RowHammer test, we write the corresponding data pattern
to the victim row ($V$ in Table~\ref{tab:data_patterns}), and to the 8 previous ($V - [1...8]$) and next ($V + [1...8]$) physically-adjacent rows. 

\begin{table}[h]
    \centering
    \footnotesize
    \caption{Data patterns used in our RowHammer {analyses.} }
    \setlength{\tabcolsep}{3pt}
    \begin{tabular}{l|cccc}
    \toprule

           \bf{Row {Address}}
           &  \bf{Colstripe{$^\dagger{}$}} & \bf{Checkered{$^\dagger{}$}} & \bf{Rowstripe{$^\dagger{}$}} & \bf{Random}\\
        \midrule
        $V^* \pm [0,2,4,6,8]$ &\verb$0x55$&\verb$0x55$ &\verb$0x00$&\verb$random$\\
        $V^* \pm [1,3,5,7]$ &\verb$0x55$&\verb$0xaa$&\verb$0xff$&\verb$random$\\
        \bottomrule
    \end{tabular}
    \begin{flushleft}
       $\quad^*V$ is the physical address of the victim row \\
       $\quad^\dagger{}${We also test the {complements} of these} patterns
    \end{flushleft}  
    \label{tab:data_patterns}
\end{table}

\vspace{0.2cm}
\noindent\textbf{Metrics.} 
We compare the \gls{ber} across all our tests at a constant hammer count of 150K per aggressor row. We also identify the DRAM cells that flip only at a particular temperature point (i.e., canary cells).

\vspace{0.2cm}
\noindent\textbf{Iterations.} To collect reliable results and estimate temperatures with \mechanism{}, we repeat every single experiment \lois{20 times} for a particular DRAM module and temperature. We use the 20 repetitions of the experiments in different ways, depending on the particular evaluation (e.g., for estimating the accuracy on absolute temperature values in Section~\ref{sec:evaluation_ber}, we use the first 10 repetitions to build the estimation model and the other 10 repetitions to estimate the model's accuracy).

\subsection{Tested DRAM Modules}

Table~\ref{table:dram_chips} summarizes the 12 DDR4 modules (120 DRAM chips) we test from four major manufacturers. With the goal of testing our hypotheses (i.e., we can spy on the temperature if we know the model of the victim DRAM, or we can reverse engineer it), we test 3 modules from each manufacturer that are exactly the same model, and have exactly the same manufacturing date. 

\begin{table}[h]
    \caption{{Summary of DDR4 DRAM chips tested.}}
    \centering
    \scriptsize{}
    \setlength\tabcolsep{2pt} 
    \begin{tabular}{lccclllll}
        \toprule
            {\bf Manufacturer} & {\bf Model{$^\dagger{}$}} & {\bf Module Id.}  & {{\bf \#Chips}}  & {{\bf Density}} & {{\bf Die}}& {\bf Org.} & {\bf Date } \\ 
            {\bf (Mfr.)} &  &   &   &  & &  & {\bf (year/week)} \\
        \midrule
        \multirow{3}{*}{A (Micron)}   & 9TBJ & 1 & 16 & 16GB & B & $\times$4 & 19/11 \\ 
          & 9TBJ & 2 & 16 & 16GB & B & $\times$4 & 19/11 \\ 
            & 9TBJ & 3 & 16 & 16GB & B & $\times$4 & 19/11 \\ 
        \midrule
       
        \multirow{3}{*}{B (Samsung)} & 8GNT & 4 & 8 & 8GB & F & $\times$8 & 21/02 \\ 
          & 8GNT & 5 & 8 & 8GB & F & $\times$8 & 21/02 \\ 
          & 8GNT & 6 & 8 & 8GB & F & $\times$8 & 21/02 \\ 
        \midrule
        \multirow{3}{*}{C (Hynix)} & S8/4 & 7 & 8 & 4GB & D & $\times$8 & 19/46 \\ 
         & S8/4 & 8 & 8 & 4GB & D & $\times$8 & 19/46 \\ 
          & S8/4 & 9 & 8 & 4GB & D & $\times$8 & 19/46 \\ 
        \midrule
        \multirow{3}{*}{D (Nanya)}   & PGRK & 10 & 8 & 8GB & C & $\times$8 & 21/12 \\ 
        & PGRK & 11 & 8 & 8GB & C & $\times$8 & 21/12 \\ 
        & PGRK & 12 & 8 & 8GB & C & $\times$8 & 21/12 \\ 
        \bottomrule
    \end{tabular}
    \begin{flushleft}
       \centering{$^\dagger{}$} Last 4 digits of the model reference.
    \end{flushleft}  
    \label{table:dram_chips}
\end{table}










\section{Fine-Grained Temperature Characterization}
\label{sec:characterization}

We characterize RowHammer under fine-grained temperature variations. We focus on the study of both \gls{ber} (Section~\ref{sec:evaluation_ber}) and canary cells (Section~\ref{sec:canary_cell_analysis}), as they are the most useful to spy on temperature (Section~\ref{sec:variantone}).

\subsection{Effect of Temperature on RowHammer BER}
\label{sec:characterization_ber}

We perform \minesh{a} fine-grained characterization of the correlation between \gls{ber} and temperature, for four different manufacturers.

\begin{figure}[ht] \centering
    \includegraphics[width=1.0\linewidth]{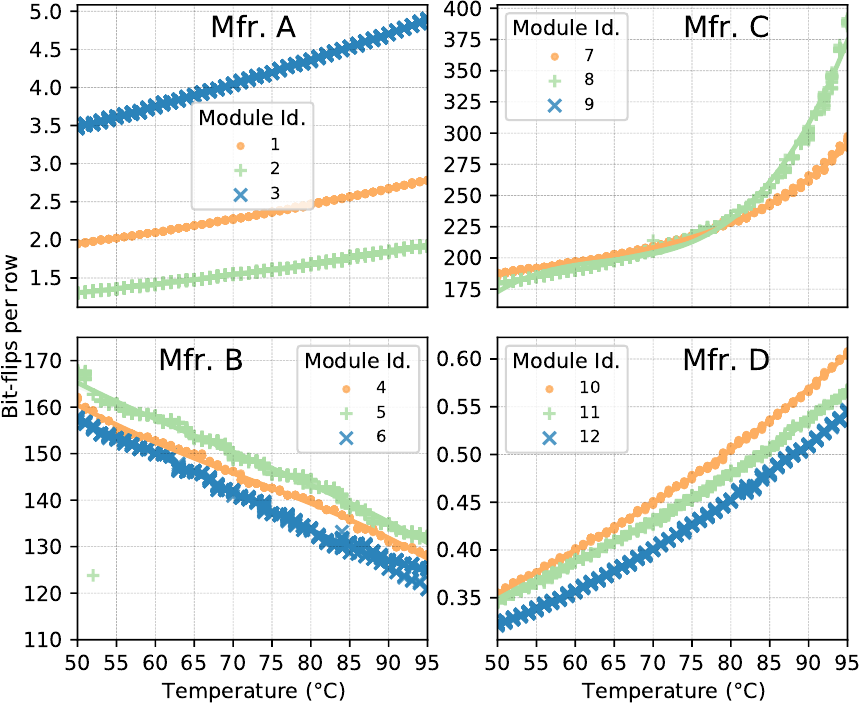}
    \caption{Correlation between RowHammer-induced bit flips per row (\gls{ber}) and temperature.} 
    \label{fig:tempBER}
\end{figure}

Figure~\ref{fig:tempBER} shows the RowHammer-induced \gls{ber} for temperatures from $50^\circ$C to $95^\circ$C, with a resolution of $1^\circ$C. We analyze the 12 DRAM modules described in Table~\ref{table:dram_chips}. We test a 192MB memory region (i.e., 24K DRAM rows) for each DRAM module, and we repeat each experiment 20 times for each tested temperature. For each module, we also plot the polynomial regression model (a continuous line with the same color as the Module's color) that better fits the BER changes across the entire temperature range.

\observation{The absolute \gls{ber} values of 2 identical modules might differ significantly.}\label{taggon:absolute_ber_same_module}

Although identical modules (i.e., same manufacturer, model, and fabrication date) might have similar \gls{ber} (e.g., modules from manufacturer B), other modules \gls{ber}'s might differ significantly. For example, module Id. 3 from Mfr. A shows on the order of 2.5$\times$ more bit flips than module Id. 2.

Thus, we can not assume that two identical modules have similar \gls{ber}. From all the DRAM modules we tested, we can only assume that  identical modules \gls{ber}'s are in the same order of magnitude.

\observation{The \gls{ber} might increase or decrease depending of the DRAM manufacturer.\label{taggon:inc_dec}}

For the same manufacturer, we always observe a consistent trend among all DRAM modules we test, which is consistent with the observations of a previous work \cite{orosa2021Deeper}. Particularly, the \gls{ber} increases with temperature in Mfrs. A, C and D, and it decreases with temperature for Mfr. B.

\observation{The \gls{ber} of two identical modules when the temperature changes follows a similar trend.}\label{taggon:ber_trend}

When we plot the polynomial regression, we observe that identical modules follow very similar curves in the plot. For example, although the absolute \gls{ber} values differ slightly, all curves from Mfr. D modules have approximately the same shape in the figure. The exception is module 8 (Mfr. C), which follows a different curve than modules 7 and 9. We speculate that Mfr. C might have used  different chips in module 8, even when the external labeling is exactly the same as in the other 2 modules.
%

\observation{Most modules show a nearly linear relation between \gls{ber} and temperature.}\label{taggon:linear_relation}

This is the case for Mfrs. A, B and D. However, module 8 from Mfr. C has a non-linear relation between \gls{ber} and temperature when the temperature is higher than \SI{70}{\celsius}. 
%


\observation{The \gls{ber} is very stable across different repetitions of the same experiment, using the same module.}\label{taggon:stable_ber}

The \gls{ber} across 20 repetitions of the experiment in one module is very stable, as we can observe in the figure by the little variation within the y-axis for each temperature.

\take{The evolution of the \gls{ber} values when increasing temperature follows a similar curve in DRAM modules with the same characteristics, even if the absolute \gls{ber} values differ significantly.\label{taggon:takeaway1}}

\vspace{0.2cm}
\noindent{\bf BER Variations in Different DRAM Regions.}
%
%
We study the \gls{ber} of 48 DRAM regions of 4MB per DRAM module, for all modules we test, for temperatures from $50^\circ$C to $95^\circ$C, with a resolution of $1^\circ$C.

\begin{figure}[ht] \centering
    \includegraphics[width=1.0\linewidth]{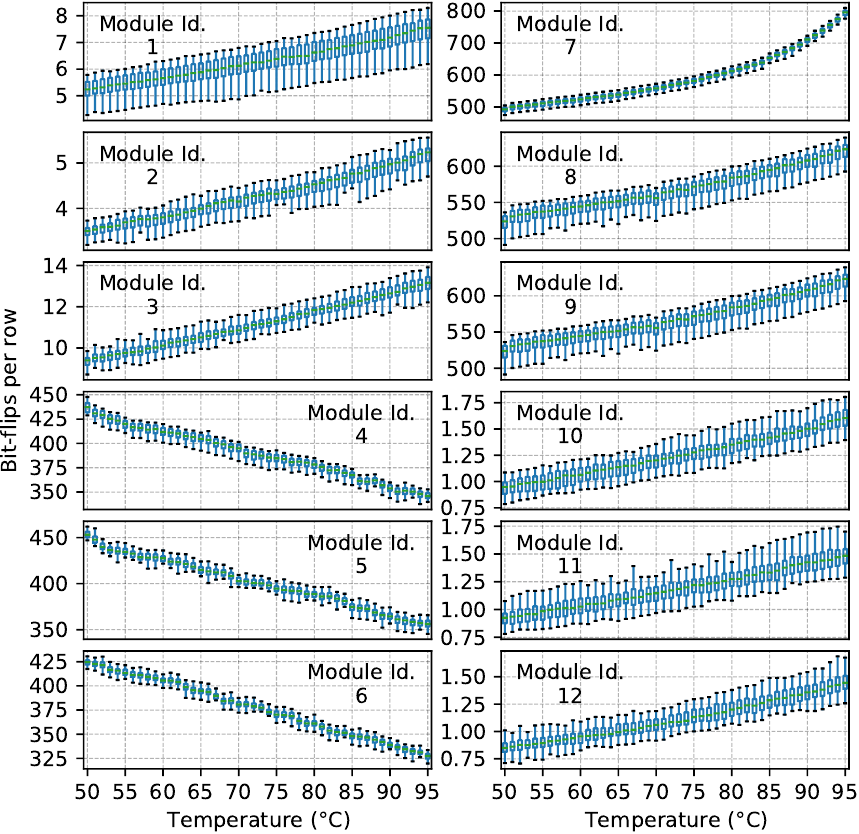}
    \caption{Box plot of RowHammer-induced \gls{ber} in 48 different 4MB memory regions from the same DRAM module.}
    \label{fig:BERreg}
\end{figure}

Figure~\ref{fig:BERreg} shows a box plot\footnote{In a box plot~\cite{Tukey1977Exploratory}, the box shows the lower and upper quartile of the data (i.e., the box spans the $25^{\mathrm{th}}$ to the $75^{\mathrm{th}}$ percentile of the data). The line in the box represents the median. The bottom and top whiskers each represent {an} additional $1.5\times$ the \emph{inter-quartile range} (IQR, the range between the bottom and the top of the box) beyond the lower and upper quartile, respectively.} that illustrates the correlation between the \gls{ber} and the temperature. For each temperature, the represented data is the \gls{ber} from the 48 different DRAM regions of the DRAM module. 

\observation{The variation of \gls{ber} values across different DRAM regions in a module is reasonably small.}\label{taggon:ber_regions}

We observe in the figure that, for all DRAM modules, the box at each temperature is pretty narrow, which indicates that the \gls{ber} is very similar for all 48 regions of each DRAM module.

\observation{The correlation of \gls{ber} with temperature follows the same trend in different regions.\label{taggon:ber_trend_regions}}

We observe that both the boxes and the whiskers follow the same trend, from which we can infer that the \gls{ber} curves from all regions are similar. We confirm this observation by comparing the curves from all 48 regions for each DRAM module (not shown in the figure).


\take{The evolution of \gls{ber} values when increasing temperature 
 follows a similar curve in different DRAM regions within the same DRAM module.\label{taggon:takeaway2}}


\vspace{-0.45cm}
\subsection{Characterization of Canary Cells}
\label{sec:canary_cell_analysis}
We characterize all DRAM modules to identify the canary cells (i.e., cells vulnerable to RowHammer at one specific temperature, but not vulnerable at any other temperature) at each temperature we \lois{test}. To identify canary cells, we select those cells that 1) experience bit flips at least once in 10 repetitions of the experiment at a particular temperature point, and 2) only experience bit flips at that temperature point.

\begin{figure}[ht] \centering
    \includegraphics[width=1.0\linewidth]{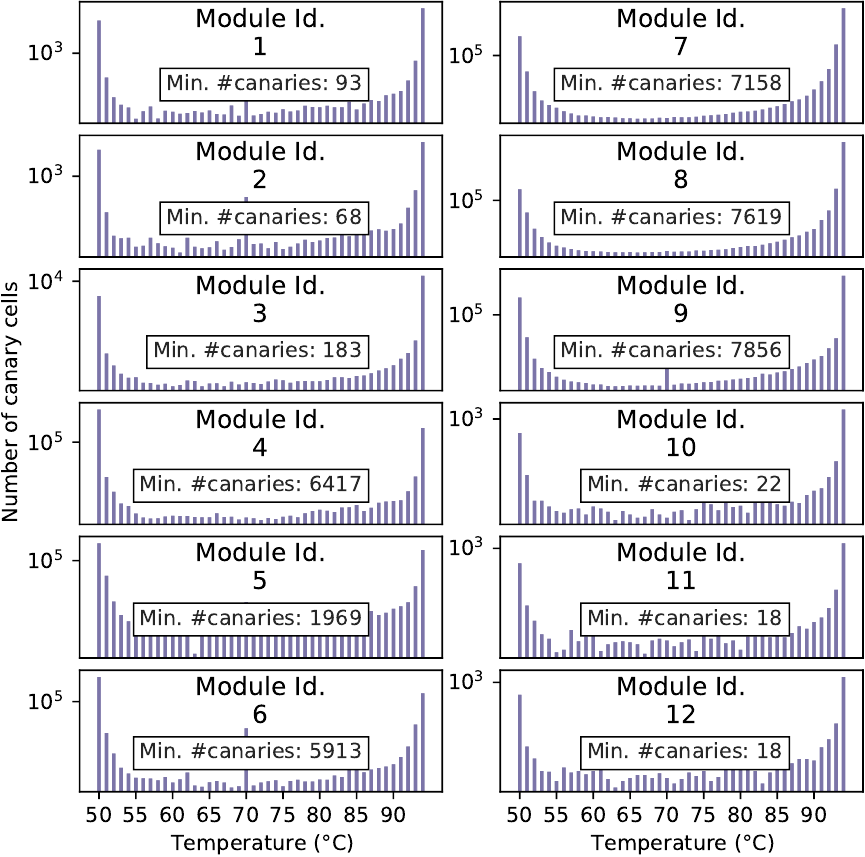}
    \caption{Number of canary DRAM cells at each temperature point.}
    \label{fig:num_canary_cells}
\end{figure}

Figure~\ref{fig:num_canary_cells} shows the number of canary cells (in logarithmic scale) per temperature point, and the minimum number of canary cells across all temperature points (Min. \#canaries). 

\observation{There are plenty of canary cells \minesh{throughout} the entire temperature range when using a high temperature resolution (\SI{1}{\celsius}). \label{taggon:canary_resolution}}

All temperature points we test have at least 18 canary cells (\minesh{module} Id. 11 and 12) in the worst case, in all modules we test. The minimum number of canary cells is large enough, \minesh{as only one canary cell is needed} to estimate the temperature at a particular temperature point.
\footnote{\minesh{Canary cells at the limits of the temperature range are more abundant than in the middle of the range. This phenomenon is caused by the limited temperature range, so some canary cells at the lower or upper limits might not be canary cells on an extended temperature range. For example, a canary cell at \SI{50}{\celsius} in a \SI{50}{\celsius}-\SI{90}{\celsius} temperature range,  might not be a canary cell in the tamperature range \SI{49}{\celsius}-\SI{90}, as the cell might also flip at both \SI{50}{\celsius} and \SI{49}{\celsius}.}}

\take{For a given temperature range, with a high temperature resolution (\SI{1}{\celsius}), there are plenty of canary cells in all modules we test.}\label{taggon:takeaway3}


\section{Overview: The SpyHammer Attack}

We describe how to perform a \mechanism{} attack that spies on \minesh{the} \emph{temperature} of the victim DRAM chip, leveraging the observations and takeaways we make in Section~\ref{sec:characterization}. Section~\ref{sec:variantone} describes how to perform a \mechanism{} attack that spies on \emph{relative temperature changes}, and Section~\ref{sec:varianttwo} describes how to perform a \mechanism{} variant that spies on \emph{absolute temperatures}.

\subsection{Spying on Relative Temperature Variations using SPYHAMMER}
\label{sec:variantone}

The basic \mechanism{} attack is based on two key observations. First, the RowHammer-induced \gls{ber} consistently increases (decreases) when the temperature increases~\cite{orosa2021Deeper} (Observation~\ref{taggon:inc_dec} in Section~\ref{sec:characterization}). We use this observation to infer if the temperature increases or decreases compared to a reference temperature point by just monitoring the \gls{ber} in a DRAM region. 
Second, the form of the curve that relates \gls{ber} and temperature is usually very similar across modules from the same manufacturer and manufacturing date (Observation~\ref{taggon:ber_trend} in Section~\ref{sec:characterization}).

Based on these observations, \mechanism{} can detect relative changes on DRAM chip temperature by 1) continuously monitoring the \gls{ber} of the victim DRAM module at different points in time, and 2) correlating the \gls{ber} changes with temperature changes on the victim DRAM module using a temperature-\gls{ber} polynomial regression model obtained from a DRAM module that has the same characteristics as the victim DRAM module. 

\vspace{0.2cm}
\noindent{\bf Canary Cell Optimization.} Obtaining consistent and reliable \gls{ber} numbers might require hammering a large region of the DRAM module, which can make the \mechanism{} attack intrusive. To solve this problem, we propose an optimization to reduce the number of hammers needed to spy on temperature. This optimization is based on the observation that some DRAM cells are vulnerable to RowHammer only at a very narrow temperature range (Observation~\ref{taggon:canary_resolution} in Section~\ref{sec:characterization}). We use this observation to make \mechanism{} faster and less intrusive by identifying the cells that flip only at one temperature point (i.e., canary cells). By using canary cells instead of \gls{ber} to spy on temperature, we can reduce the number of DRAM accesses to perform the \mechanism{} attack.


The canary cell optimization is performed in two steps. First, the attacker enrolls canary cells for different temperatures by associating vulnerable DRAM cells to a particular \gls{ber} value (i.e., to a particular temperature). For an accurate enrollment phase, the victim DRAM experiments should \agy{operate} at different temperatures, so the \gls{ber} and the vulnerable DRAM cells also change. For enrolling as many canary cells as possible, it is important that the victim DRAM module experiences all possible temperature variations in regular operational conditions. Depending on the particular system to monitor and the required accuracy of the attack, a complete and full enrollment process might take from a few hours to one year (e.g., the system might experience different temperatures at different seasons of the year). Second, after the canary cells are identified, the attacker can monitor these canary cells to spy \agy{on} the temperature of the victim DRAM module.


\vspace{0.2cm}
\noindent{\bf Threat Model.}
We assume an attacker with \agy{\emph{no}} prior knowledge or physical access to the victim DRAM module. Thus, the victim's software and hardware can not be characterized or modified by the attacker prior to the attack. 
With the goal of building models that correlate \gls{ber} with temperature, the attacker can purchase many DRAM modules from different manufacturers and has the infrastructure to control the DRAM temperature of those modules in a fine-grained manner. To build an accurate model, the attacker needs to characterize a DRAM module very similar to the victim module (e.g., \agy{the} same manufacturer and model), which requires knowing or reverse-engineering the manufacturer and model of the victim DRAM module remotely.


\subsection{Spying on Absolute Temperature Values using SPYHAMMER}
\label{sec:varianttwo}

This variant of the \mechanism{} attack aims to spy on absolute temperature values. To this end, the attacker must have physical access to the DRAM module and characterize it under different temperature conditions \agy{before} the attack, using a similar methodology as in Section~\ref{sec:variantone}, but without the need to reverse-engineer the DRAM model. Because the attacker uses the victim DRAM module to build the polynomial regression model, and they have local and accurate control over the DRAM temperature, the attack is more precise, as we demonstrate in Section~\ref{sec:evaluation_ber}. The rest of the attack is identical to the attack in Section~\ref{sec:variantone}. We discuss the limitations of this attack model in Section~\ref{sec:limitations}.

\vspace{0.2cm}
\noindent{\bf Threat Model.}
\label{sec:threatmodel2}
We assume an attacker that has access to the victim DRAM module prior to performing the attack \agy{and} can characterize the bit errors caused by RowHammer in a controlled environment at different temperatures.

\section{The Six Steps of the SpyHammer Attack}

This section describes in detail the six steps of the \mechanism{} attack on relative temperature changes (Section~\ref{sec:variantone}):
\begin{enumerate}[label=\arabic*)]
    \item Identify the victim DRAM \lois{module}
    \item Build a polynomial regression model from a DRAM module that is the same model as the victim
    \item Allocate a contiguous DRAM memory region in the victim system
    \item Monitor the \gls{ber} of the victim DRAM module continuously
    \item Enroll canary cells in the victim DRAM module
    \item Monitor canary cells in the victim DRAM module to infer its temperature
\end{enumerate}

\subsection{Step 1: Identify the Victim DRAM Module}
\label{sec:identify_victim}

\mechanism{} requires reverse engineering the manufacturer and model of the victim DRAM module. Our methodology identifies the DRAM manufacturer first and then identifies the technology node of the DRAM module. There are several techniques to reverse-engineer the victim DRAM manufacturer and model remotely, without physical access or modification to the victim system. Next, we explain some of these techniques an attacker can use based on new observations and observations from previous works.

We classify our methodologies into two categories: 1)~methodologies that are very simple and quick to implement but that can lead to certain inaccuracies (Section~\ref{sec:identify_quickly}), and 2)~methodologies that are more complex but that generally provide more accurate results (Section~\ref{sec:identify_accurate}). The best methodology for each case can be chosen based on the required accuracy or attack time.

\subsubsection{Quickly Identifying the DRAM Manufacturer and Model}
\label{sec:identify_quickly}
We can quickly infer some characteristics of the DRAM manufacturer of the victim DRAM module remotely by using two simple techniques based on RowHammer. These techniques can not accurately distinguish between Mfr. A and Mfr. D modules, so they might have to be complemented with more sophisticated techniques (Section~\ref{sec:identify_accurate}) when required.

\vspace{0.2cm}
\noindent{\bf Mfr. B: Unique Logical-to-Physical Row Mapping.} DRAM manufacturers use DRAM internal mapping schemes to translate
memory-controller-visible row addresses to physical row addresses in DRAM~\cite{kim2014flipping, smith1981laser, horiguchi1997redundancy, keeth2001dram, itoh2013vlsi, liu2013experimental,seshadri2015gather, khan2016parbor, khan2017detecting, lee2017design, tatar2018defeating, barenghi2018software, cojocar2020rowhammer,  patel2020beer, yaglikci2021blockhammer}.
We use the logical-to-physical row mapping to uniquely identify Mfr. B modules. We observe that Mfr.~A, Mfr.~C, and Mfr.~D modules use a sequential logical-to-physical row mapping, whereas Mfr. B uses a unique logical-to-physical row mapping in which two adjacent row addresses sent from the memory controller might map to non-adjacent physical locations in the DRAM chip. 

To discover the logical-to-physical row mapping of a DRAM module, we use the observation that, when performing a single-sided RowHammer attack, the rows with more RowHammer-induced \agy{bitflips} are the rows that are physically adjacent to the attacker row. Using this observation, we perform a simple iterative algorithm that infers the logical-to-physical mapping. Table~\ref{tab:logical_to_physical_mapping} shows the logical-to-physical row mapping from all four major DRAM manufacturers we test. In the table, each element in the address, physical ($phy$) or logical ($log$), represents a bit, where bit 0 is the least significant bit of the address.

\setcounter{table}{2}
\begin{table}[h]
  \begin{center}
  \footnotesize
  \begin{tabular}{l|c} 
  \textbf{Logical-to-Physical Row Mapping} & \textbf{Manufacturer (Mfr.)} \\ 
   \toprule
   $phy[x] = log[x]{^\dagger{}}$ & A (Micron), C (Hynix),  \\
   & and D (Nanya)\\
   \midrule
    $phy[0] = log[0]$ & \multirow{4}{*}{B (Samsung)}\\ 
    $phy[1] = log[3] \oplus log[1]$ & \\
    $phy[2] = log[2] \oplus log[3]$ &\\
    $phy[y] = log[y]{^\ddagger{}}$ & \\
\bottomrule
\end{tabular}
    \begin{flushleft}
       \centering{$^\dagger{}$} where x ranges from 0 to N.\\
       \centering{$^\ddagger{}$} where y ranges from 3 to N.
    \end{flushleft}  
\caption{Logical-to-physical mapping of DRAM row addresses.}
\label{tab:logical_to_physical_mapping}
\end{center}
\end{table}

We make two observations. First, Mfr. B  is the only manufacturer that uses a non-sequential logical-to-physical row mapping. Second, all Mfr. B modules have the same logical-to-physical row mapping in all modules we test. We verify that Mfr. B uses this mapping also for modules other than the ones we use for performing our thorough characterization (Table~\ref{table:dram_chips}). We conclude that an attacker can identify Mfr. B modules \minesh{solely} by reverse-engineering their unique logical-to-physical row mapping.

\vspace{0.2cm}
\noindent{\bf Mfr. A and Mfr. D: Single-sided RowHammer Tests.} We make the new observation that performing single-sided RowHammer usually only affects one neighboring row in Mfr. A and Mfr. D modules. For example, hammering row $X$ causes bitflips only in $X+1$, and hammering row $X+1$ causes bitflips only in row $X$. In Mfr. B and Mfr. C modules, when performing a single-sided RowHammer attack, two victim rows are susceptible to \agy{bitflips}. We speculate that this observation is caused by microarchitectural design decisions.\footnote{DRAM manufacturers do not reveal any detail about the internal DRAM microarchitecture} implementation. We conclude that we can use a single-sided RowHammer attack to identify DRAM modules from Mfr. A and D.

\vspace{0.2cm}
\noindent{\bf Identifying the DRAM Model.}
Modules from the same manufacturer might have different characteristics, depending on the manufacturing process, manufacturing date, etc. To identify the particular DRAM model of a DRAM module, we use the observation that the \gls{ber} of modules from the same manufacturer but different technology nodes differ significantly. This observation could also be used to infer the DRAM manufacturer, as the absolute \gls{ber} values from different manufacturers also differ very significantly in the modules we test (see Section~\ref{sec:evaluation_ber}).

\subsubsection{Accurately Identifying the DRAM Manufacturer and Module}
\label{sec:identify_accurate}
U-TRR~\cite{hassan2021uncovering} assesses the security guarantees of recent DRAM chips by reverse engineering proprietary on-die RowHammer mitigation mechanisms, commonly known as Target Row Refresh (TRR). TRR detects and refreshes potential RowHammer-victim rows, but its exact implementations are not openly disclosed. U-TRR is based on the new observation that data retention failures in DRAM enable a side channel that leaks information on how TRR refreshes potential victim rows.
The authors show in their evaluation that it is possible to reverse engineer the TRR mechanism for many different DRAM models from 3 major manufacturers. Our observation is that each manufacturer implements TRR in a different way, and, for modules from the same manufacturer, there are also a wide variety of TRR implementations. We conclude that we can use the techniques proposed by \cite{hassan2021uncovering} to uniquely identify a particular DRAM model. 

\subsection{Step 2: Build the Polynomial Regression Model}
\label{sec:step3}

Although the absolute \gls{ber} of two different modules from the same manufacturer and technology node might differ significantly,  \gls{ber} and temperature are very correlated (i.e., the shape of the polynomial regression curve is very similar in most cases). Thus, using the data obtained from the \gls{ber} characterization, an attacker can 1)~build a polynomial regression model from a DRAM module with the same characteristics as the victim DRAM module and 2)~use the model to estimate the relative temperature changes of the victim DRAM module.

Using a polynomial regression model, we can estimate the temperature change. In our evaluation, we demonstrate that a polynomial regression model of order 3 is enough to estimate temperature with $\pm$\SI{5}{\celsius} accuracy when estimating relative temperature changes, and $\pm$\SI{1}{\celsius} accuracy when estimating absolute temperatures, for most DRAM modules we test.

\subsection{Step 3: Allocate a Contiguous Memory Region}
Unlike other RowHammer attacks~\cite{razavi2016flip}, \mechanism{} does not require sophisticated techniques to place the victim memory row (not controlled by the attacker) into a particular physical row that is adjacent to the aggressor row controlled by the attacker. Instead, in \mechanism{}, both the victim and the aggressor row are in the attacker's own memory space. 

The attacker only has two requirements when allocating memory from the victim system. First, the memory region should be as contiguous as possible (i.e., the memory space is not very fragmented), so when performing hammers, the aggressor and victim row are neighbors with high probability. This is important not only for maximising the \gls{ber}, but also for not corrupting the memory of other processes that use memory in adjacent memory regions. Second, for using the canary cell optimization, the memory region should not migrate to different physical locations during the attack, as the canary cells change from region to region. When using \gls{ber} to spy on temperature, this is not required, as different regions within the same DRAM module have similar \gls{ber} (see Section~\ref{sec:evaluation_ber}).

To identify if the physical memory region changes at different points in time, the attacker can use memory deduplication to reverse-map any physical page into a virtual page~\cite{razavi2016flip}.

\subsection{Step 4: Monitor the BER of the Victim DRAM Module}
\label{sec:step4}

\mechanism{} is carried out in a memory region entirely in the attacker's address space.  To characterize the \gls{ber} of the memory region, the attacker simply needs to perform a RowHammer attack to every row within their address space with a double-sided RowHammer attack (as described in Section~\ref{sec:methodology}), and count the total number of RowHammer-induced \agy{bitflips} in the memory region. More sophisticated attacks can be used to trigger the attack (e.g., Blacksmith~\cite{jattke_blacksmith_2022}), but we evaluate a double sided attack to simplify the comparison and to extract more clear conclusions.

The attacker initially establishes an estimated reference temperature by correlating the measured \gls{ber} to the estimated temperature using the polynomial regression model (Step~3). In subsequent \gls{ber} measurements and temperature estimations, the attacker can estimate the relative temperature changes using the model.

\subsection{Step 5: Enroll Canary Cells}
\label{sec:step5}

To improve performance and reduce the probability of the \mechanism{} attack \agy{being} detected, we identify and monitor canary cells, i.e., cells that only flip at a particular temperature point. By using canary cells, an attacker only needs to characterize (i.e., hammer) a few cells in the memory region instead of the entire memory region. Before using canary cells to estimate temperature, the attacker needs to identify those canary cells in a variety of different temperatures (i.e., enrollment). Enrollment requires characterizing the victim system during a \minesh{long-enough} time period to the system work in a wide variety of temperatures.

The canary enrollment process registers the cells that will be used as canaries for detecting temperature changes. This process requires monitoring DRAM cells while characterizing the \gls{ber} (Step 4). To this end, the attacker 1) associates every different \gls{ber} value (with a specified error tolerance) with the DRAM cells that experience \agy{bitflips} at only at the associated \gls{ber} (i.e., canary cells), and 2) eliminates duplicated canary cells from the previously generated \gls{ber}-canary cells pairs. By using this methodology iteratively, the attacker can identify which cells are vulnerable to RowHammer at a specific \gls{ber} value.
%


\subsection{Step 6: Monitor Canary Cells to Infer the Temperature of the victim's DRAM}
\label{sec:step6}

After enrolling the canary cells, the attacker monitors canary cells (i.e., check if the enrolled canary cells are vulnerable to RowHammer) to detect temperature changes, instead of accessing large regions of DRAM to calculate the \gls{ber}. 
By comparing the \gls{ber} associated with different canary cells, the attacker can infer relative temperature changes using the same methodology as used in Step~4.

To minimize the number of DRAM accesses, the attacker can use a limited \agy{number of canary cells per temperature or access DRAM rows with many} canary cells. 

\section{SpyHammer Evaluation}
\label{sec:evaluation1}

We evaluate \minesh{\mechanism{}'s} accuracy and sensitivities when using both \gls{ber} (Section~\ref{sec:evaluation_ber}) and canary cells (Section~\ref{sec:canary_cell_analysis}) to spy on relative temperature changes and absolute temperature.

\subsection{BER Analysis}
\label{sec:evaluation_ber}

From Observations \minesh{\ref{taggon:absolute_ber_same_module}-\ref{taggon:stable_ber}} and Takeaway \ref{taggon:takeaway1} (Section \ref{sec:characterization_ber}), we infer that the similar correlation between \gls{ber} and temperature between modules from the same manufacturer can be used to estimate the temperature of the victim DRAM module.\footnote{For a particular model of a Mfr. D module (not shown in our evaluation results), we can not observe enough RowHammer-induced bitflips to make conclusive observations about the relation between RowHammer and Temperature.}
To do so, we find that a polynomial regression model of order 3 is sufficient to accurately model the correlation between \gls{ber} and temperature for all modules we test. Table~\ref{tab:linear_regresion} shows the polynomial regression equations inferred for each DRAM module. 

\begin{table}[ht]
 \scriptsize\setlength{\tabcolsep}{2pt}
  \begin{center}
  \begin{tabular}{c|l} 
  \toprule
  {\bf Id.}  & {\bf Polynomial Regression}\\
  \midrule
  1 &   $y = -(1.9*10^{-6})*x^3 + (4.8*10^{-4})*x^2 - (2.2*10^{-2})*x + 2.1 $\\
  2 &   $y = +(6.8*10^{-7})*x^3 - (6.4*10^{-5})*x^2 + (1.2*10^{-2})*x + 0.8$ \\
  3 &   $y = +(4.0*10^{-7})*x^3 +(3.0*10^{-5})*x^2  + (2.0*10^{-2})*x + 2.3$ \\
  \midrule
  4 &   $y = - (1.9*10^{-4})*x^3 + (4.0*10^{-2})*x^2 - 3.5*x +260$\\
  5 &   $y = +(9.5*10^{-5})*x^3 - (2.1*10^{-2})*x^2 + 0.8*x +157.7$ \\
  6 & $y =  +(2.4*10^{-4})*x^3 -(4.9*10^{-2})*x^2 + (2.5)*x + 121.8$ \\
  \midrule
  7 &   $y = + (1.4*10^{-3})*x^3 -0.2*x^2 + 15.6*x -152.2 $\\
  8 &   $y = +(5.1*10^{-3})*x^3 -1.0*x^2 +64.1*x -1201.8 $\\
  9 &   $y = +(8.1*10^{-5})*x^3 - (9.9*10^{-3})*x^2 + (0.9)*x + 167.6$\\
  \midrule
  10 &   $y = +(4.5*10^{-7})*x^3 - (6.3*10^{-5})*x^2 + (7.4*10^{-3})*x + 0.1$ \\
  11 &   $y =  +(3.4*10^{-7})*x^3 - (4.3*10^{-5})*x^2 + (5.7*10^{-3})*x + 0.1$\\ 
  12 &   $y = - (1.2*10^{-7})*x^3 + (6.4*10^{-5})*x^2 + (2.5*10^{-3})*x + 0.3) $\\
\bottomrule
\end{tabular}
\caption{Polynomial regression that models the correlation between \gls{ber} and temperature.}
\label{tab:linear_regresion}
\end{center}
\end{table}

We observe that the regression curve is similar between modules from the same manufacturer. We conclude that we can spy on temperature by using a polynomial regression model from a DRAM module that is the same model as the victim DRAM module. However, the accuracy of this methodology can significantly change depending on the DRAM module and threat model.

\vspace{0.2cm}
\noindent{\bf Methodology for Measuring Accuracy.} To evaluate the accuracy of our methodology to estimate temperature using a polynomial regression model, we select a sequence of 720 random temperature points in the tested range (i.e., from \SI{50}{\celsius} to \SI{95}{\celsius}), and we measure the accuracy of the estimation for all temperatures in the sequence. For spying on relative temperature changes, we use the change between consecutive temperature points in the sequence, and for spying on absolute temperature, we use each temperature point in the sequence. We make sure that the sequence of random temperatures includes extreme temperature changes (e.g., from \SI{95}{\celsius} to \SI{50}{\celsius}), and small temperature changes (e.g., from \SI{50}{\celsius} to \SI{51}{\celsius}).

\begin{figure}[ht] \centering
    \includegraphics[width=1.0\linewidth]{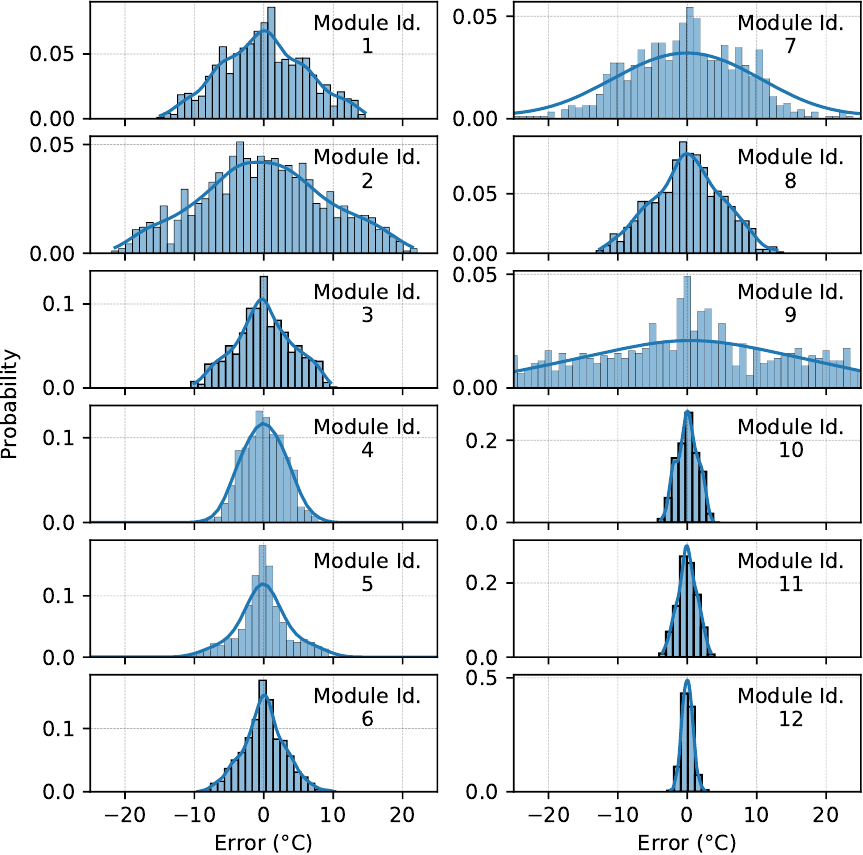}
\caption{Error of the relative temperature change estimation obtained using a polynomial regression model from a DRAM module that is the same model as the victim DRAM module.}
    \label{fig:relative_temperatureResolution}
\end{figure}

\vspace{0.2cm}
\noindent{\bf Accuracy when Spying on Relative Temperature Changes.} We study the accuracy of spying the relative temperature changes of the victim DRAM module using a polynomial regression model obtained from a DRAM module that is the same model as the victim DRAM module. Figure~\ref{fig:relative_temperatureResolution} shows the probability distribution of the temperature error when we use the polynomial regression model. For each victim DRAM module of each manufacturer, we use a polynomial regression model obtained from one of the other two modules we test from the same manufacturer. 

Table~\ref{tab:percentile_relative} shows the maximum error in the estimation of the relative temperature changes (in \SI{}{\celsius}), for $90^{\mathrm{th}}$ percentile of the error distribution. We show values for temperature changes up to \SI{45}{\celsius} (L), and for temperature changes up to \SI{5}{\celsius} (S), which should be the common case if the \gls{ber} monitoring frequency is large enough.

\begin{table}[ht]
 \setlength{\tabcolsep}{2pt}
  \begin{center}
  \footnotesize
  \begin{tabular}{c|c|c|c|c|c|c|c|c|c|c|c|c} 
  \toprule
  {\bf Module Id.} & {\bf 1} & {\bf 2} & {\bf 3} & {\bf 4} & {\bf 5} & {\bf 6} & {\bf 7} & {\bf 8} & {\bf 9} & {\bf 10} & {\bf 11} & {\bf 12}\\
    \midrule
  {\bf Error (L)$^\dagger{}$ (\SI{}{\celsius})} & 7.8 & 12.6 & 5.6 & 3.9 & 4.6 & 3.9 & 14.5 & 6.6 & 23.1 & 2.0 & 1.9 & 1.0 \\
  {\bf Error (S)$^\ddagger{}$ (\SI{}{\celsius})} & 3.1 & 3.1 & 3.3 & 3.4 & 2.4 & 3.2 & 2.6 & 3.3 & 2.6 & 2.0 & 1.9 & 1.0\\
\bottomrule
\end{tabular}
    \begin{flushleft}
       \centering{$^\dagger{}$} for relative temperature changes up to \SI{45}{\celsius}.\\
       \centering{$^\ddagger{}$} for relative temperature changes up to \SI{5}{\celsius}
    \end{flushleft}  
\caption{Maximum temperature error (in \SI{}{\celsius}) for $90^{\mathrm{th}}$ percentile of the error distribution, when estimating relative temperature changes.}
\label{tab:percentile_relative}
\end{center}
\end{table}

We make two observations. First, the error of the temperature estimated by the regression model \lois{is} reasonably low for modules from Mfr. B (Module Id. 4, 5, 6) and Mfr. D (Module Id. 10, 11, 12), because the correlation between \gls{ber} and temperature is very similar across all modules from the same manufacturer. We find that the error is larger for Mfr. A (Module Id. 1, 2, 3) and Mfr. C (Module Id. 7, 8, 9) because in each of these manufacturers, there is one module that has a slightly different curve than the other two modules. Second, in many cases, the estimation error is zero (i.e., error = \SI{0}{\celsius}). We observe that when the temperature change is small (e.g., $<$\SI{5}{\celsius}), the estimation error is small for all modules ($<$\SI{3.5}{\celsius}), whereas when the temperature change is large (e.g., up to \SI{45}{\celsius}) the estimation error is much larger (only 3 modules have less than \SI{3.5}{\celsius} error). 

We make two conclusions. First, for all possible temperature changes (i.e., from very small to very large temperature changes), a polynomial regression model works well if the modules show very similar correlations between \gls{ber} and temperature. Second, for small temperature changes, we can get accurate temperature estimations from all modules we test. By sampling at a high enough frequency, this should be always the case. 

\begin{figure}[ht] \centering
    \includegraphics[width=1.0\linewidth]{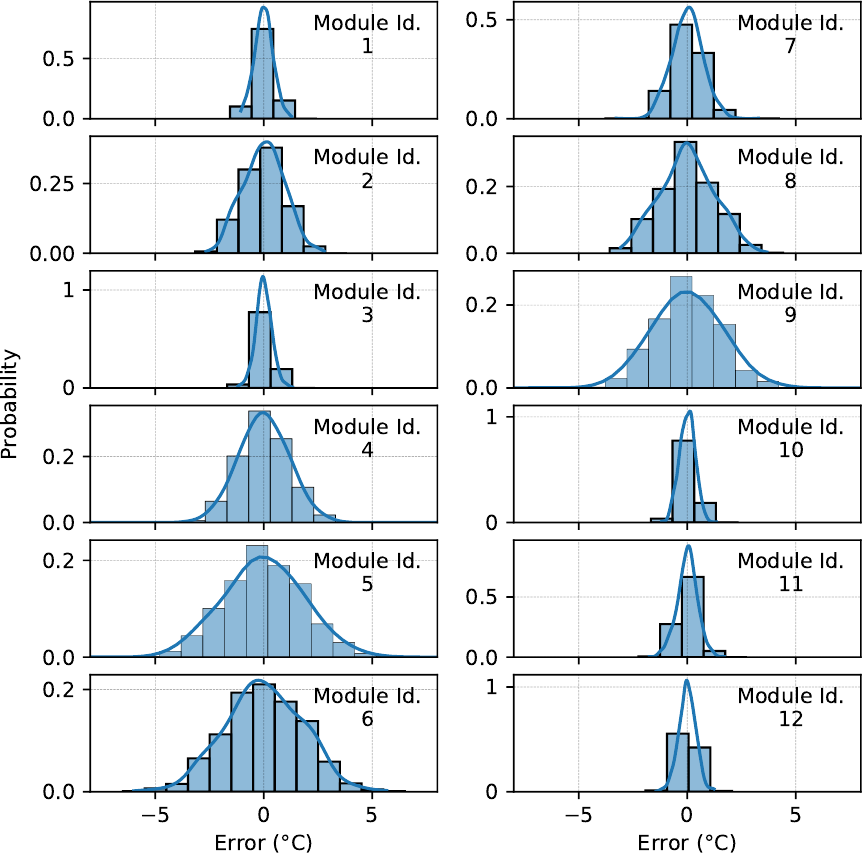}
    \caption{\lois{Error of the absolute temperature estimation obtained with a polynomial regression model from the victim DRAM module}.} 
    \label{fig:absolute_temperatureResolution}
\end{figure}

\vspace{0.2cm}
\noindent{\bf Accuracy when Spying on Absolute Temperature Values.} We study the accuracy of spying on the absolute temperature of the victim DRAM module using a polynomial regression model obtained from the victim DRAM module itself. Figure~\ref{fig:absolute_temperatureResolution} shows the probability distribution of the error when estimating the absolute temperature.

Table~\ref{tab:percentile_absolute} shows the maximum error in the estimation of the absolute temperatures (in \SI{}{\celsius}), for $90^{\mathrm{th}}$ percentile of the error distribution.

\begin{table}[ht]
 \footnotesize
 \setlength{\tabcolsep}{2pt}
  \begin{center}
  \begin{tabular}{c|c|c|c|c|c|c|c|c|c|c|c|c} 
  \toprule
  {\bf Module Id.} & {\bf 1} & {\bf 2} & {\bf 3} & {\bf 4} & {\bf 5} & {\bf 6} & {\bf 7} & {\bf 8} & {\bf 9} & {\bf 10} & {\bf 11} & {\bf 12}\\
    \midrule
  {\bf Error (\SI{}{\celsius})} & 0.5 & 1.2 & 0.5 & 1.4 & 2.3 & 2.3 & 0.9 & 1.7 & 1.9 & 0.5 & 0.6 & 0.5 \\
\bottomrule
\end{tabular}

\caption{Maximum temperature error (in \SI{}{\celsius}) for $90^{\mathrm{th}}$ percentile of the error distribution, when estimating absolute temperatures.}
\label{tab:percentile_absolute}
\end{center}
\end{table}

We make the main observation that the absolute temperature estimations are very accurate for most of the DRAM modules. All modules we tested show an error lower than \SI{2.5}{\celsius} for $90^{\mathrm{th}}$ percentile \lois{of the error distribution}. We conclude that 1) using a regression model from the victim DRAM module provides accurate results, and 2) a polynomial regression model is enough to accurately estimate absolute temperatures in most modules.

\vspace{0.2cm}
\noindent{\bf BER Variations in Different DRAM Regions.}
We conclude that the attacker can use any region in memory to perform the attack, which simplifies the requirements of the attack (i.e., no need for reverse engineering the physical location of the memory region to find an appropriate region). 

\begin{figure}[ht] \centering
    \includegraphics[width=1.0\linewidth]{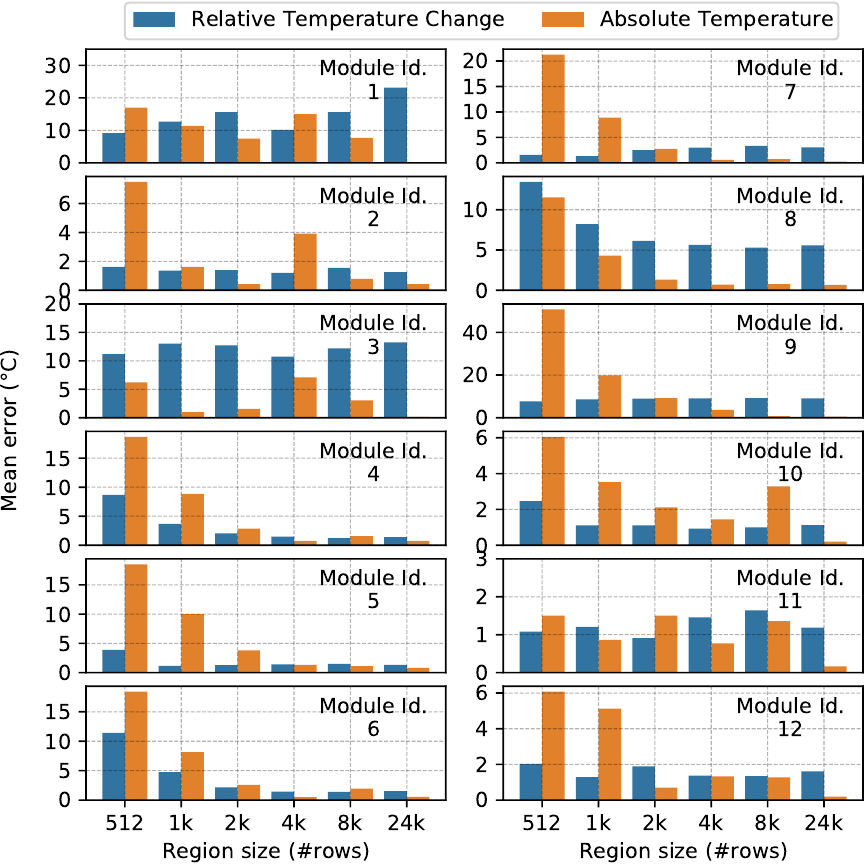}
    \caption{Mean error of the temperature estimated by the polynomial regression model for different region sizes.}
    \label{fig:regionSize}
\end{figure}

\vspace{0.2cm}
\noindent{\bf Sensitivity to Different Region Sizes.}
To improve performance and reduce the probability of the attack being detected, the attacker can reduce the size of the DRAM region used for performing the attack. Figure~\ref{fig:regionSize} shows the mean error of the temperature change estimated by the polynomial regression model for different region sizes from 512 rows (4MB) to 24k rows (192MB). The color of the bars represents the two \lois{evaluated threat models}: 1) spying on relative temperature changes, and 2) spying on absolute temperatures.

We make three observations. 
First, we can reduce the region size from 24k to 2k with minor to no increase in the mean error for all cases we test. 
Second, for region sizes lower than 2k, the mean error increases significantly in many cases. For example, modules from Mfr. B (Module Id. 4, 5 6) show a very significant error increase in the absolute temperature estimation.
 Third, the mean error values show similar trends for both relative temperature changes and absolute temperature, for most modules we test. However, the error increases more quickly when reducing the region size when estimating absolute temperatures.

We conclude that the attacker can use regions of size as small as 2k rows (i.e., 16MB) for spying on temperature, while maintaining reasonable levels of accuracy.

\subsection{Canary Cell Analysis}
\label{sec:canary_cell_eval}

From Obserbation  \minesh{\ref{taggon:canary_resolution}} and Takeaway \ref{taggon:takeaway3} (Section \ref{sec:canary_cell_analysis}), we infer that all modules we test have a large enough number of canary cells (i.e., more than one) for each temperature point, so they can be used to reliably infer the DRAM temperature.

\begin{figure}[ht] \centering
    \includegraphics[width=1.0\linewidth]{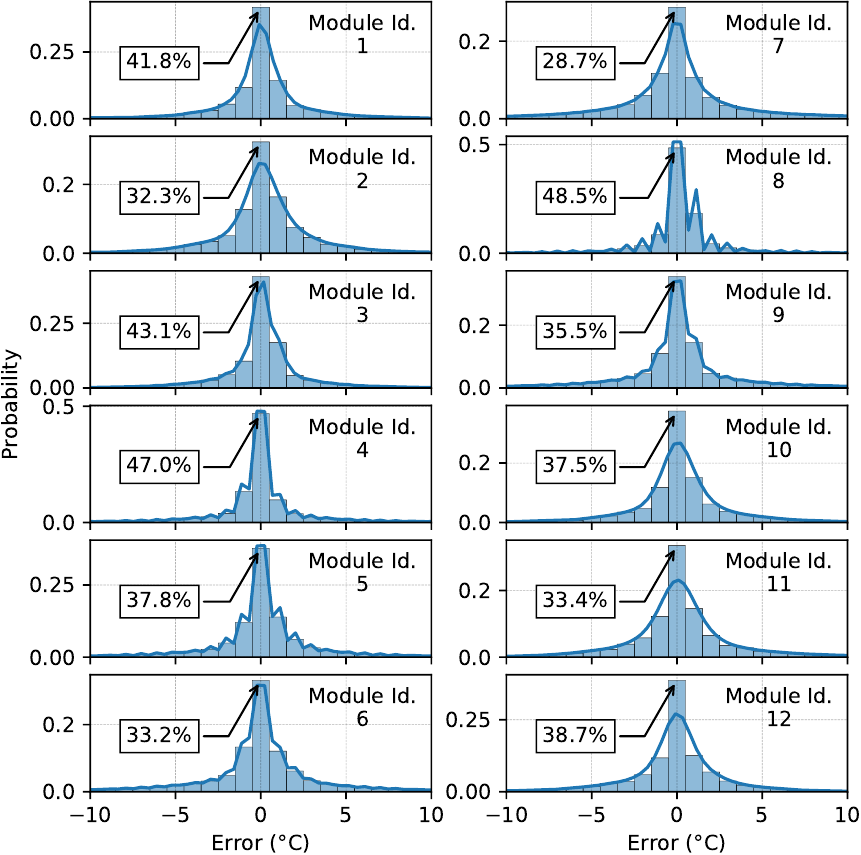}
    \caption{Probability distribution of the temperature estimation errors using canary cells.}
    \label{fig:temperatureResolutioncanary}
\end{figure}

\vspace{0.2cm}
\noindent{\bf Canary Cell Accuracy.}
\label{sec:evaluation_canarycell}
We measure the accuracy of \mechanism{} when using canary cells.   Figure~\ref{fig:temperatureResolutioncanary} shows the probability distribution of the temperature errors when spying on absolute temperature using canary cells.\footnote{Similarly, canary cells can be also used to estimate relative temperature changes.} We use 10 iterations of our experiments to perform the enrollment process (i.e., identify the canary cells), and another 10 different iterations of our experiments to monitor the canary cells and spy on temperature. 

We make two main observations. First, canary cells can estimate the temperature accurately (i.e., no errors) with more than 25\% probability, with some cases close to 50\% (e.g., Module Id. 4, 8). Second, canary cells can be used to estimate temperature with less than $\pm$\SI{5}{\celsius} error with very high probability. 

We conclude that canary cells can be used to estimate temperature with significant accuracy while reducing the number of hammers significantly. For example, if an attacker wants to know if the temperature is \SI{60}{\celsius}, it only has to monitor a canary cell that flips only at that temperature point, instead of hammering a large region of memory (see Section~\ref{sec:evaluation_ber}).

\section{Discussion and Limitations}
\label{sec:discussion}

\subsection{Discussion}
\label{sec:discussion1}

\mechanism{} is the first practical attack that spies on DRAM temperature without any software or hardware modification to the victim system. As many prior works demonstrate, DRAM devices are becoming \agy{increasingly} vulnerable to RowHammer~\cite{kim2020revisiting,orosa2021Deeper}, and they can be remotely induced at system level~\cite{gruss2016rowhammer,tatar2018throwhammer}, without physical access, which makes \mechanism{} challenging to mitigate.

Unlike other RowHammer attacks~\cite{gruss2016rowhammer,tatar2018throwhammer,razavi2016flip,van2016drammer}, \mechanism{} is much simpler to perform because of two main reasons. First, \mechanism{} happens entirely in the attacker address space, not requiring to trigger bitflips in other processes, \lois{which makes the attack difficult to detect}. Second, the attacker does not require performing complex memory templates, and it does \emph{not} depend on the memory allocator of the operating system~\cite{razavi2016flip}. As we demonstrate in Section~\ref{sec:evaluation_ber}, the correlation between \gls{ber} and temperature is very similar across different regions of the same DRAM module, thus \mechanism{} also does not need to understand the exact physical location of the allocated memory.

For these reasons, we believe that \mechanism{} is a simple practical attack that can compromise the security and privacy of any system that uses modern DRAM modules.





\subsection{Limitations}
\label{sec:limitations}

\mechanism{} has one main limitation. To spy on absolute temperature changes, the attacker \agy{needs} to characterize the victim DRAM module \agy{before} performing the attack, in an environment controlled by the attacker with precise temperature control. Without previous access to the victim DRAM device, we can only infer relative temperature changes over time. This is also the case for any mechanism proposed by \lois{a} previous work~\cite{xiong2019spying}. The reason of this limitation is that each DRAM chip presents unpredictable variation caused by process variation, \lois{even if we compare it with other modules from the same manufacturer and technology node}. For example, we find that the \gls{ber} of two modules from the same manufacturer and technology node can differ up to a factor of 2.5$\times$ (see Figure~\ref{fig:tempBER}).


\section{Countermeasures}

\mechanism{} can spy on DRAM temperature without any modification to the victim system. The correlation between RowHammer-induded errors and temperature is inherent to the DRAM device. There are two types of countermeasures against \mechanism{}.

First, general RowHammer defense mechanisms that prevent agains RowHammer bit flips, independently of the temperature. A good RowHammer defense mechanism that can mitigate most RowHammer bit flips would be also effective to prevent \mechanism{}. 
%
There is an emerging body of work that provides efficient RowHammer defense mechanisms~\cite{AppleRefInc, kim2014flipping, kim2014architectural, aichinger2015ddr, bains2015row, aweke2016anvil, bains2016distributed, bains2016row, gomez2016dummy, brasser2017can, son2017making, konoth2018zebram, seyedzadeh2018cbt, van2018guardion, hassan2019crow, lee2019twice, kang2020cattwo, park2020graphene, yaglikci2021blockhammer, yaglikci2021security, devaux2021method, you2019mrloc, jedec2017ddr4, jedec2015hbm, jedec2020ddr5, jedec2020lpddr5,greenfield2012throttling,woo2023scalable,juffinger2023csi,wi2023shadow,marazzi2023rega,zhou2023p,di2023copy,saileshwar2022randomized,qureshi2022hydra,kim2022mithril,fakhrzadehgan2022safeguard,saxena2022aqua,joardar2022machine,marazzi2022protrr,loughlin2022moesi,bostanci2024comet,olgun2024abacus} that can \lois{also} be used to mitigate \mechanism{}. 
For example, BlockHammer~\cite{yaglikci2021blockhammer} selectively throttles memory accesses that could otherwise potentially cause RowHammer bitflips. The key idea of BlockHammer is to 1) track row activation rates using area-efficient Bloom filters, and 2) use the tracking data to ensure that no row is ever activated rapidly enough to induce RowHammer bitflips.

Second, specific RowHammer defense mechanisms that obfuscate the relation between \gls{ber} and temperature by introducing a temperature-dependent parameters in the mechanism. 

Thus, we conclude that \mechanism{} should be mitigated with general and effective RowHammer mitigation mechanisms. To date, existing DRAM modules still do not employ such effective techniques \onur{that are proven to be secure~\cite{hassan2021uncovering,orosa2021Deeper,kim2020revisiting,jattke_blacksmith_2022,frigo2020trrespass}}.


\section{Related Work}
\label{sec:related_work}

\minesh{\mechanism{}} is the first attack that can remotely spy on temperature without compromising the victim system (i.e., no software or hardware modification to the victim system). 

\vspace{0.2cm}
\noindent{\bf Attacks that Spy on Temperature.} Xiong et al.~\cite{xiong2019spying} proposes an attack that spies on DRAM temperature by using the observation that when the temperature increases~\cite{liu2013experimental}, the leakage of the DRAM cells also increases, thus retention errors are manifested earlier. The main limitation of this attack is that, to observe bit errors caused by retention failures, the cell needs to be discharged for many seconds. This work has two main limitations. First, the technique requires disabling DRAM refreshes in the victim system, which can corrupt the data in critical data structures of the operating system, which would break the system. Second, disabling refreshes requires modifications to the victim system, which in turn requires the attacker to have physical access to the device before conducting the attack.
Compared to Xiong et al.~\cite{xiong2019spying}, \mechanism{} can spy on temperature without any hardware or software modifications to the victim system. Also, \mechanism{} does not require to disable refreshes, which enables remotely \minesh{spying} on temperature without corrupting the data of other processes running on the same system.

\vspace{0.2cm}
\noindent{\bf Other RowHammer Attacks.} Many prior works exploit RowHammer to perform system-level attacks~\cite{gruss2018another,gruss2016rowhammer,hong2019terminal,ji2019pinpoint,kwong2020rambleed,lipp2018nethammer,razavi2016flip,seaborn2015exploiting,van2016drammer,xiao2016one,yao2020deephammer,jattke_blacksmith_2022,cohen2022hammerscope,tobah2022spechammer,fahr2022frodo,mus2023jolt,zhang2022implicit,kogler2022half,kangsledgehammer,tobahgo}. These attacks can perform denial of service~\cite{gruss2018another,lipp2018nethammer}, privilege escalation~\cite{gruss2018another,lipp2018nethammer,gruss2016rowhammer,ji2019pinpoint,razavi2016flip,seaborn2015exploiting,van2016drammer,xiao2016one,jattke_blacksmith_2022,zhang2022implicit}, secret data leakage~\cite{kwong2020rambleed,cohen2022hammerscope,tobah2022spechammer,tobahgo}, manipulation of the application correctness~\cite{hong2019terminal,yao2020deephammer} or recovering private keys~\cite{fahr2022frodo,mus2023jolt}. Compared to these existing works, \mechanism{} is a much less intrusive \lois{attack}, as it does not involve manipulating data from any other process other than the attacker process.

\vspace{0.2cm}
\noindent{\bf Characterization of Real DRAM Chips.} There are several works that extensively characterize RowHammer using real DRAM chips~\cite{kim2014flipping,kim2020revisiting,orosa2021Deeper,yauglikcci2022understanding,olgun2023experimental,lang2023blaster,he2023whistleblower,yaglikci2024svard,yuksel2024functionallycomplete}. The first RowHammer work~\cite{kim2014flipping} that that investigates the vulnerability in detail for the first time 1) analyzes 129 commodity DDR3 DRAM modules, 2) characterizes the sensitivity of RowHammer to refresh rate, activation rate, and the physical distance between aggressor and victim rows, and 3) analyzes several potential solutions. The second extensive RowHammer characterization~\cite{kim2020revisiting}, conducted in 2020, analyzes RowHammer scalability by performing experiments on 1580 DDR3, DDR4, and LPDDR4 commodity DRAM chips from different DRAM generations and technology nodes, demonstrating that RowHammer has become more problematic over time. The third work~\cite{orosa2021Deeper}, conducted in 2021, studies the sensitivity of RowHammer to DRAM chip temperature, aggressor row active time, and victim DRAM cell's physical location, by performing experiments on 248 DDR4 and 24 DDR3 modern DRAM chips from four major manufacturers.

Even though these works rigorously characterize various RowHammer aspects, they do not analyze the effects of temperature in great detail: 1) they do not perform fine-grained analyses (e.g., \SI{1}{\celsius} steps), 2) they do not empirically analyze their observations from the attacker perspective in great detail, and 3) they do not build and experimentally demonstrate new RowHammer attacks based on the correlation between RowHammer vulnerability and temperature.

\vspace{0.2cm}
\noindent{\bf PUFs and Physical Cryptography.}
\minesh{Our work also relates to the recent areas of physical unclonable functions (PUFs) and physical cryptography. In these areas, the intrinsic, physical characteristics of hardware are employed to either enable new cryptographic and security schemes, or to launch new classes of attacks.  Optical PUFs and digital silicon PUFs have been pioneered early on by Pappu et al.\ and Gassend et al. \ \cite{pappu2002physical,gassend2002silicon}, and have been advanced in a large number of follow-up works \cite{suh2007physical,lugli2013physical,merli2011side,ruhrmair2013power,lugli2013physical,csaba2010application,chen2009analog,ruhrmair2012simpl,van2014protocol,ruhrmair2011physical,sauer2017sensitized,chen2011circuit,gao2018efficient,horstmeyer2015physically,guajardo2007fpga,majzoobi2009techniques,majzoobi2010fpga,kumar2008butterfly,ruhrmair2022secret,csaba2009chip,jaeger2010random,ruhrmair2010towards,ruehrmair2012method,ruhrmair2013practical,ruhrmair2012practical}.
Also, DRAM PUFs and Rowhammer PUFs have been \lois{proposed} recently \cite{kim2018dram,tehranipoor2015dram,sutar2016d,schaller2017intrinsic,anagnostopoulos2018intrinsic}.}



\section{Conclusion}

Recent studies demonstrate that new DRAM devices are becoming increasingly more vulnerable to RowHammer, and many works demonstrate system-level attacks for privilege escalation or information leakage. In this work, we provide the first analysis of RowHammer under fine-grained temperature variations, yielding nine key observations. We leverage our empirical observations to spy on DRAM temperature. 
We build a new RowHammer attack, called \mechanism{}, that spies on the temperature of critical systems such as industrial production lines, \lois{self-driving} or semi-automated vehicles, and medical systems, without any modification to the victim system. We propose two variants of \mechanism{} for two different threat models. First, if the attacker cannot characterize the victim DRAM module before the attack, they can use \mechanism{} to spy on \emph{relative temperature changes} on the victim system. Second, if the attacker can characterize the victim DRAM module before the attack, they can use \mechanism{} to spy on \emph{absolute temperature}. 
Our evaluation shows that \mechanism{} can 1) spy on relative temperature changes with an error of $\pm$\SI{3.5}{\celsius}, and 2) spy on absolute temperature changes with an error of $\pm$\SI{2.5}{\celsius}, for all 12 DRAM modules from \onur{four} manufacturers, at the $90^{\mathrm{th}}$ percentile of tested temperature points. 

We conclude that \mechanism{} is a simple and effective attack that can spy on temperature of critical systems with no modifications or prior knowledge about the victim system. We believe that \mechanism{} can be a potential threat to the security and privacy of systems until a definitive and completely-secure RowHammer defense mechanism is adopted, which is a large challenge given that RowHammer vulnerability continues to worsen with technology scaling.


\section*{Acknowledgments}
We thank the SAFARI Research Group members for {useful} feedback and the stimulating intellectual environment they provide. We acknowledge the generous gifts provided by our industrial partners{, including} Google, Huawei, Intel, Microsoft, and VMware, and support from the Microsoft Swiss Joint Research Center. Ulrich Rühmair acknowledges support by the Air Force Office of Scientific Research (AFOSR) via grant FA9550-21-1-0039.

\balance
\bibliographystyle{IEEEtran}
\bibliography{refs}



\end{document}